# Measurement and Analysis of Radiofrequency Radiation Exposure Level from Different Mobile Base Transceiver Stations in Ajaokuta and Environs, Nigeria

Ushie, P. O.[a,1], Victor U. J. Nwankwo[b,2], Ayinmode Bolaji[c] and Osahun, O. D.[d]

[a] *Department of Physics, Cross River University of Technology, Calabar, Nigeria*
[b] *Department of Physics and Energy Studies, Salem University, Lokoja, Nigeria*
[c] *Department of Physics, University of Ibadan, Ibadan, Nigeria*
[d] *Department of physics, University of Benin, Benin City, Nigeria*

***Abstract:*** *We present the result of a preliminary assessment of radio-frequency radiation exposure from selected mobile base stations in Ajaokuta environs. The Power density of RF radiation within a radial distance of 125m was measured. Although values fluctuated due to the influence of other factors, including wave interference from other electromagnetic sources around reference base stations, we show from analysis that radiation exposure level is below the standard limit (4.5W/m$^2$ for 900MHz and 9W/m$^2$ for 18000MHz) set by the International Commission on Non-ionizing Radiation Protection (ICNIRP) and other regulatory agencies.*

## I. Introduction

The term radio frequency (RF) covering all frequencies used for communication, radar, satellites etc. is a rate of oscillation in the range of about 3 kHz to 300 GHz, which corresponds to the frequency of radio waves, and the alternating currents which carry radio signals. RF usually refers to electrical rather than mechanical oscillations, perhaps since RF cannot be seen it has been confused with ionizing radiation. The consequence of exposure to both can be stated as been very different. Radio transmissions relatively a very new technology, which had its beginning in the theoretical work of Maxwell and experimental work of Hertz, a German physicist. Many others also made contributions, including the development of devices which could detect the presence of radio waves (Aslan, 1972). Over the years, the transmission of radio waves has become an established technology which is taken for granted and which among other things provides for broadcasting to our homes for entertainment, the most recent development resulting in the domestic satellite dish antenna (Kitchen, 2001).

In the case of radio transmitters, however, the whole intention is to transmit RF energy into free space and the antenna used to do so is specifically designed to achieve this objective. Very low frequencies, e.g. mains power frequency do not give rise to any significant amount of radiation. However, as we increase the frequency, then it becomes increasingly possible to radiate electromagnetic waves, giving a suitable antenna to act as an efficient launcher. Wavelength ($\lambda$) is an important parameter in considering antenna system and propagation since it is a factor in determining the physical dimension of antennas. It relate to the velocity (c) and frequencies ($f$) of radiation as;

$$c = \lambda f \qquad (1)$$

Radio waves can therefore be referred to either by the wavelength or the frequency. Radiations are of two types; ionizing and non-ionizing radiation. Ionizing radiation is radiation capable of ejecting electrons from atoms and molecules with the resultant production of harmful free radicals. There is a minimum quantum energy below which this disruption cannot take place. Since the human body is largely water, the water molecule is used to define this minimum level (Hopfer, 1980). The highest RF used as standard for RF safety is 300GHz which corresponds to a wavelength of 10-300 meters and lies in the extra high frequency (EHF) band of the radio frequency spectrum (Gregory, 2000). However, in radio transmitters using very high supply of voltages, ionizing radiation is in the form of X-rays. It is therefore clear that ionizing radiation is not inherent in the RF energy but rather that both forms of radiation can co-exist to produce hazards that we must be conscious of. Non-ionizing radiation cannot ionize matter because its energy is lower than the ionization potential of atoms or molecules of the absorber. The term non-ionizing radiation thus refers to all types of electromagnetic radiation that do not carry enough energy per quantum to ionize atoms or molecules of the absorber. Near ultraviolet radiation, visible light, infrared photons, microwaves, and radio waves are examples of non-ionizing radiation (Guy, 1987).

---

[1] Corresponding author: pat205ushie@yahoo.com
[2] TWAS-Bose Research Fellow, S.N. Bose National Centre for Basic Sciences, Kolkata, India.

## II. Materials And Method

A handheld three-axis RF meter (electrosmog meter) was used for the measurement of the power density of electromagnetic radiation from the mobile base stations. The meter is a Broad band device for monitoring high frequency radiation in the range of 50 *MHz* to 3.5 *GHz*. It is used in three-axis (isotropic) measurement mode and includes audio alarm with adjustable threshold and 200 point manual memory function, extremely sensitive. A 4-digit LCD display offers mV/m, A/m, W/cm$^2$ or W/m$^2$. Display resolution: 0.1V/m, 0.1µW/m$^2$, 0.001µW/cm$^2$, and 0.001µW/m$^2$. The meter measures the value Electric Field, **E** and converts it into Magnetic Field, **H** and the power density *S* (i.e the power per unit area) expressed in Watts per Meter squared (W/m$^2$).

Most of the base stations have at least 3 sectorial antennas on it, which means that they cover 360° sector area (Ayinmode, 2010; Victor et al, 2012). This implies that measurement can be taken in any convenient direction around the station. Measurements were taken every 25m to 150m radius from each mast. On getting to the site, a quick environmental assessment was made. Such as the density of houses around, vegetation and the visibility of any other mast are assessed.

The meter was set to the triaxial measurement mode and also to the maximum instantaneous measurement mode, to measure the maximum instantaneous power density at each point. Each measurement was made by holding the meter away from the body, at arm's length and at about 1.5m above the sea level pointing towards the mast as suggested by Ismail et al (2010). The values of the measured power densities were taken and recoded when the meter is stable (about 3 mins). Precautions were taken as much as possible, so that the measured value were not influenced by other sources. Movement of the meter while measurements were taken was avoided to reduce excessive field strength values due to electrostatic charges.

## III. Results And Discussion

Table 1: Measured values of Power density

| | | Power density in µW/m$^2$ | | | | | | | | |
|---|---|---|---|---|---|---|---|---|---|---|
| S/N | Distance (m) | ST1 Etisalat | ST2 Etisalat | ST3 Glo | ST4 MTN | ST5 Glo | ST6 MTN | ST7 Glo | ST8 MTN | ST9 Airtel | ST10 Glo |
| 1 | 0 | 608.7 | 384.0 | 420.3 | 152.2 | 131.1 | 206.1 | 122.1 | 274.2 | 112.2 | 332.1 |
| 2 | 25 | 806.7 | 954.2 | 509.3 | 171.5 | 157.1 | 268.2 | 132.3 | 300.0 | 128.9 | 378.0 |
| 3 | 50 | 435.2 | 242.4 | 551.5 | 261.1 | 150.8 | 534.2 | 15.9 | 311.3 | 150.5 | 390.1 |
| 4 | 75 | 195.9 | 461.4 | 225.3 | 1250.0 | 378.4 | 1730.0 | 2.2 | 1180.0 | 164.4 | 249.5 |
| 5 | 100 | 500.7 | 673.7 | 224.0 | 145.9 | 88.0 | 341.2 | 5.4 | 549.1 | 72.1 | 120.1 |
| 6 | 125 | 240.0 | 393.1 | 272.8 | 80.5 | 49.4 | 82.8 | 1.1 | 82.2 | 289.9 | 7.9 |

*ST—station

**Etisalat Base Transceiver Station (BTS)**

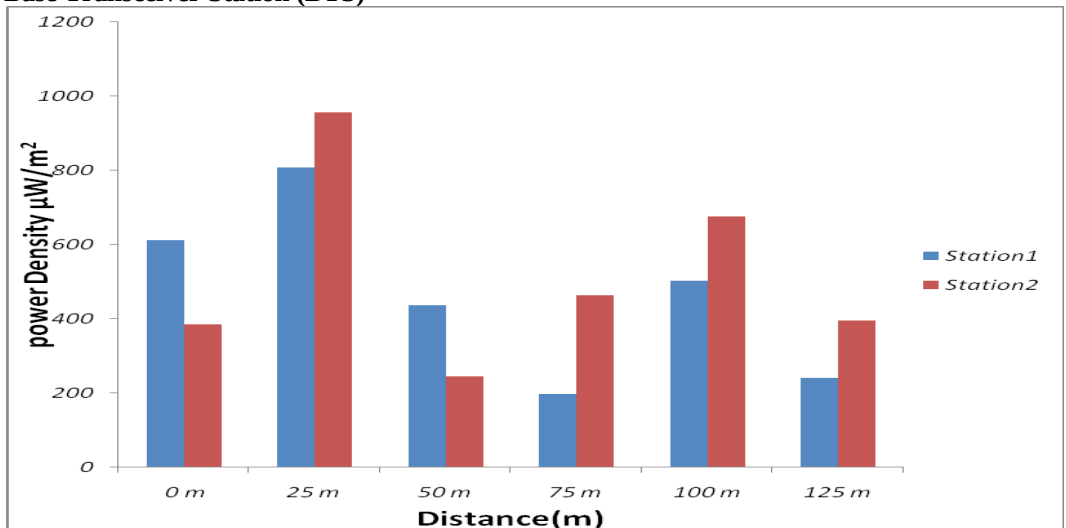

Figure 1: Plot of power density (µW/m$^2$) against distance (m).

**Glo Base Transceiver Station (BTS)**

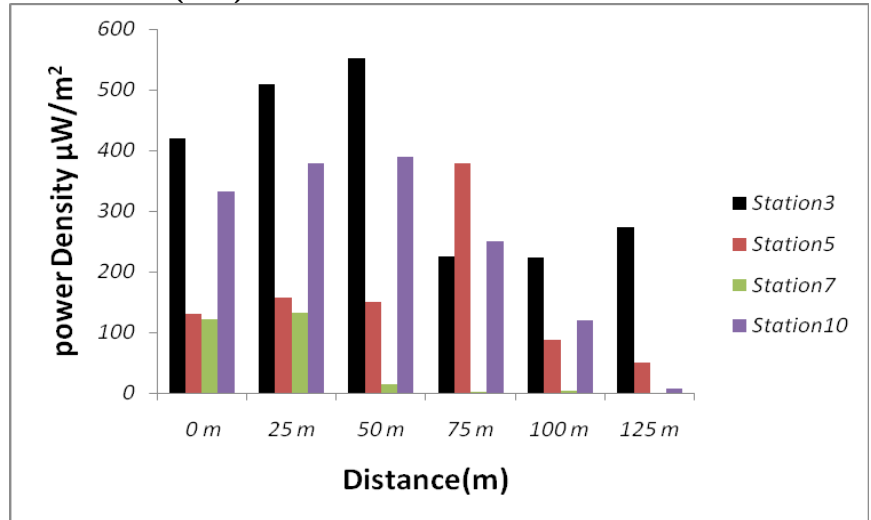

Figure 2: Plot of power density (μW/m$^2$) against distance (m)

**MTN Base Transceiver Station (BTS)**

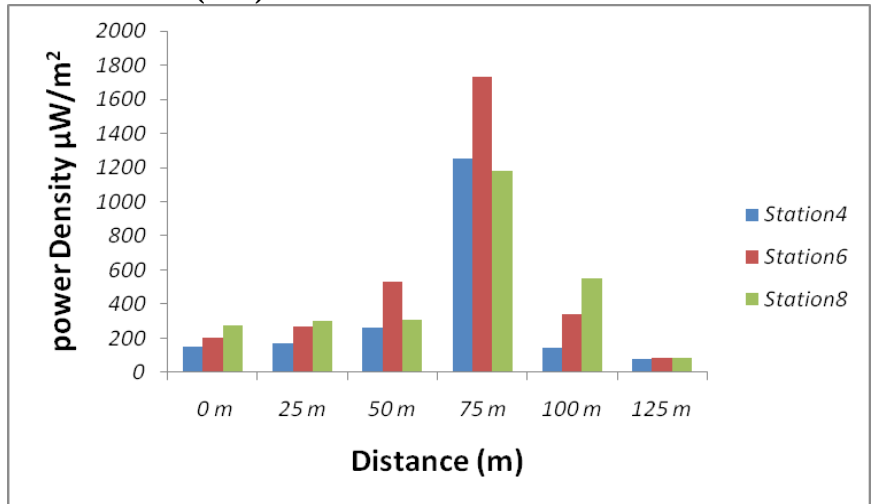

Figure 3: Plot of power density (μW/m$^2$) against distance (m)

**Airtel-BTS**

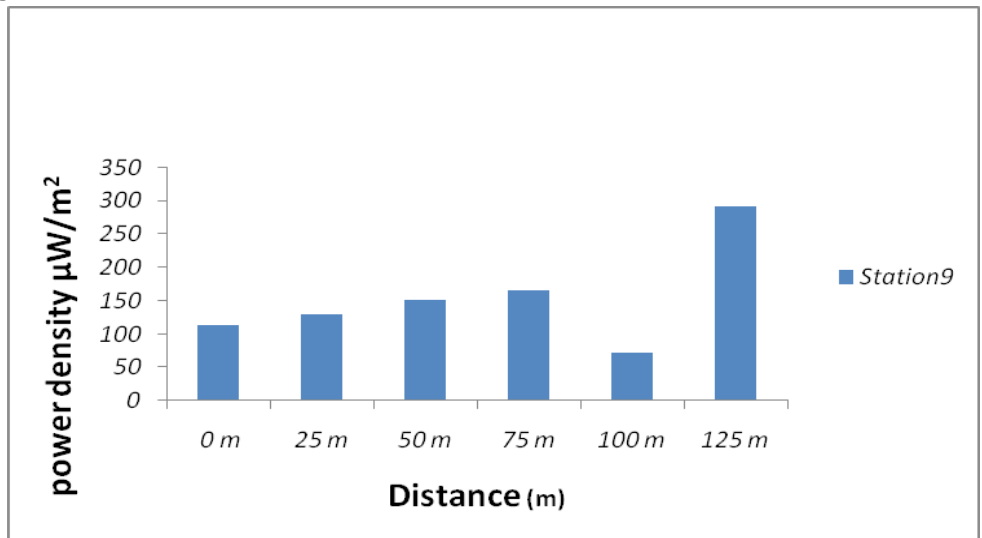

Figure 4: Plot of power density (μW/m$^2$) against distance (m)

**Plot Summary:**

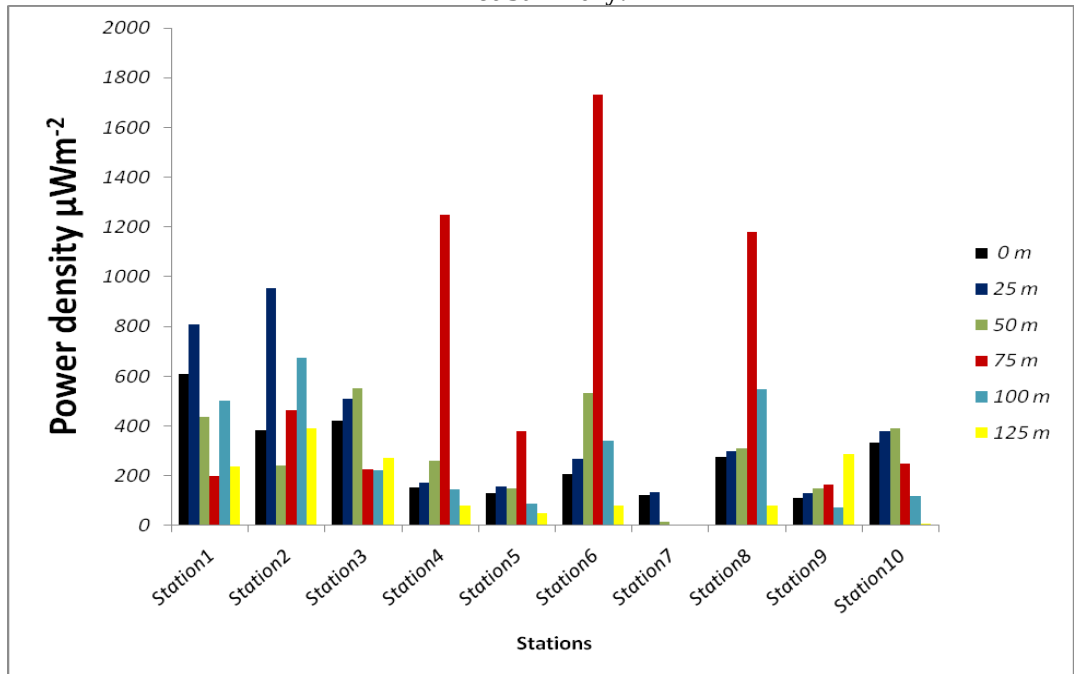

Figure 5: Plot of power density (μW/m$^2$) against stations

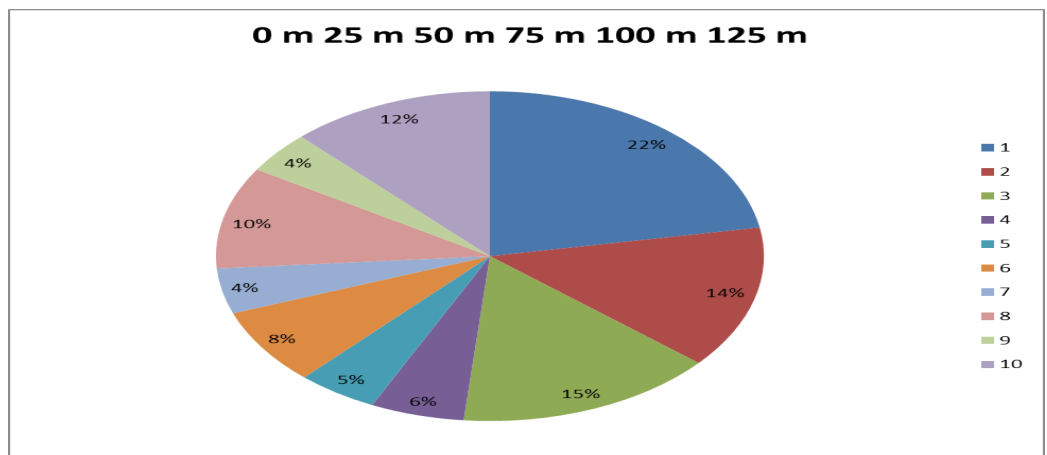

Figure 6: Percentage contribution of each station

### IV. Discussion

The highest value of power density was recorded at 25m radius of an Etisalat BTS - Base Transceiver Station (fig 1). The value dropped and then increased again at 100m from the foot of the base stations. Though this increase might be influenced by other RF emission gadgets like radio transmitters, TV antennas etc. Globacom BTS (fig 2) recorded the highest radio frequency emission at 50m from the base station and dropped as the distance from the foot of the base stations increases. MTN BTS (fig 3) registered its highest emission at 75m from the foot of the base stations. Airtel BTS (fig 4) recorded its highest radio frequency emission at 125m from the base station (this may however due to additional influence from other sources/base stations around this reference base station). Values were minimal and below the ICNIRP set standard. In all cases, the radio frequency radiation emissions from the base stations are below $10^7 \mu Wm^{-2}$, the standard limit set by the international commission on non-ionizing radiation protection (ICNIRP). However, values were assumed to be influenced by some known and unknown factors such as EM radiation from FM and TV antennas, satellite dish etc.

## V. Conclusion

The study started with an introduction to the concept of radio frequency radiation. It also highlighted various sources of the emission. Measurement of the Power density from ten (10) base transceiver stations (BTS) were measured covering about 4 networks – Etisalat, Globacom, MTN and Airtel. Although measurements were influenced by unavoidable factors, values were found to be below the standard limit (4.5W/m$^2$ for 900MHz and 9W/m$^2$ for 18000MHz) set by the International Commission on Non-ionizing Radiation Protection (ICNIRP) and other regulatory agencies.

## Acknowledgements

We are grateful to the Department of Physics and Energy Studies, Salem University, Lokoja and Dr. Norbert N. Jibiri, of Radiation and Health Physics Laboratory, Department of Physics, University of Ibadan Nigeria for their helpful contributions during this work.